\documentclass[aps]{revtex4}
\usepackage[english]{babel}
\usepackage{epsfig,float,subfigure,amssymb,graphicx}
\newcommand{\beq}{\begin{equation}}
\newcommand{\eeq}{\end{equation}}
\newcommand{\bea}{\begin{eqnarray}}
\newcommand{\eea}{\end{eqnarray}}
\newcommand{\meio}{{}^1\!/{}_{\!2}}
\newcommand{\tmeio}{{}^3\!/{}_{\!2}}

\newcommand{\tquarto}{{}^3\!/{}_{\!4}}
\newcommand{\quarto}{{}^1\!/{}_{\!4}}
\begin{document}
%\baselinestretch{1cm}
\title{Analytic results in the position-dependent mass Schrodinger problem}
\author{M. S. Cunha${}^{1}$\footnote{E-mail: marcony.cunha@uece.br} and H. R. Christiansen${}^{2}$\footnote{E-mail: hugo.christiansen@uece.br}}
\affiliation{${}^{1}$Grupo de F\'{\i}sica Te\'orica, State University of Cear\'a (UECE), Av. Paranjana 1700, 60740-000 Fortaleza - CE, Brazil}
\affiliation{${}^{2}$State University Vale do Acara\'{u}, Av. da Universidade 850, 62040-370 Sobral - CE, Brazil}
%
%\date{Received: date / Accepted: date}
%
%
%\maketitle%
\begin{abstract}
We investigate the Schrodinger equation for a particle with a nonuniform solitonic mass density.
First, we discuss in extent the (nontrivial) position-dependent mass
 $V(x)=0$ case whose solutions are hypergeometric functions in $\tanh^2 x$.
Then, we consider an external hyperbolic-tangent potential.
We show that the effective quantum mechanical problem is given by a Heun class equation and
find analytically an eigenbasis for the space of solutions.
We also compute the eigenstates for a potential of the form $V(x)=V_0 \sinh^2x$.
\end{abstract}

\maketitle
\section{Introduction}
The study of the Schrodinger equation with a position-dependent mass (PDM) has been a
matter of interest since the early days of Solid State physics.
Indeed, many of the important questions in the theory of solids concern the
non-relativistic motion of electrons in periodic
lattices perturbed by the effect of impurities, as it happens in typical semiconductors \cite{slater}.

The theoretical understanding of transport phenomena in semiconductors of a
position-dependent chemical composition is  at the root of this problem. The PDM idea arises
after the effect of the periodic field on the electrons. In this context, the electron mass was originally
replaced by a mass tensor
whose elements were determined by the unperturbed band structure \cite{luttingerkohn}.
To be more specific, the electronic wave packet near the top or the bottom of an energy band is the quantum mechanical entity related to the effective PDM concept.
As explained by von Roos \cite{vonroos}, the Wannier-Slater theorem \cite{slater,wannier}
in its simplest form
states that the envelope function $F(r, t)$ (for the conduction band, for instance) obeys a
Schrodinger-like  equation $[E(k)(-i\nabla)+ V(r)]F(r, t) =i\hbar \dot F(r, t)$
where $E(k)$ represents the conduction band energy of
the unperturbed crystal, $k$ is the crystal momentum and $V(r)$ is the potential
due to external sources (e.g. electromagnetic fields) or superficial impurities.
In simple cases, by starting with the one-electron approximation
of the many-body Hamiltonian one can approximate this equation
with a position-dependent effective mass Schrodinger equation for the envelope
\beq %
\left[-\frac{\hbar^2}{2 m(r)}\nabla^2 + E(0)+ V(r)\right]F(r, t) =i\hbar \dot F(r,t).\label{first}
\eeq
In nonuniform semiconductors, with a position-dependent chemical composition, extensions
of the theorem require not only modifications of the potential term but also of the kinetic operator.
In fact, it is already evident in Eq. (\ref{first}) where the kinetic term
is manifestly non-hermitian. A natural improvement to fix this problem is
the use of a symmetrized operator \cite{gora,bastard}
%\beq T= -\frac{\hbar^2}{4}\left(\frac{1}{m(r)}\nabla^2+\nabla^2\frac{1}{m(r)}\right)\eeq
\beq \hat T= \frac{1}{4}\left(\frac{1}{m(r)}\hat p^2+\hat p^2\frac{1}{m(r)}\right)\eeq
but neither this solves completely the question \cite{shewell} so several other orderings were
postulated in subsequent years
\cite{vonroos,BDD,tdlee,zhukroemer,likhun,thomsenetc,young,einevol1,einevol2,levyleblon}.

In the last decade, the present subject has gained renewed interest for both mathematical
and phenomenological reasons. Issues such as hermiticity, operator ordering and solubility
together with application and uses of the PDM Schrodinger equation related to  {molecular and atomic
physics} \cite{willatzenlassen2007,chinos200407,severtazcan2008,quesne2008,mustafa2009,midyaroy2009,ardasever2011,
hamdouni2011,mustafa2011,midya2011,sinha2011,sinha2012}
as well as to supersymmetry and relativistic problems \cite{plastino99,alhaidari2004,tanaka2006,
mustafa2008,midyaetal} have been
matter of intense activity in the fields of quantum mechanics and condensed matter physics
in the last years.
%
%increasing relevance in describing the motion of electrons in problems of compositionally
% graded crystals [M. R. Geller and W. Kohn, Phys. Rev. Lett. 70, 3103 (1993).]
% quantum dots [Ll. Serra and E. Lipparini, Europhys. Lett. 40, 667 (1997).]
% liquid crystals [M. Barranco, M. Pi, S. M. Gatica, E. S. Hernandez and J. Navarro, Phys. Rev. B56,
%8997 (1997)]. The appearance of PDM is also well known in the energy density
%functional approach to the nuclear many-body problem [P. Ring and P. Schuck, The Nuclear
% Many-Body Problem (Springer-Verlag, 1980)] and its applications in the context of nonlocal terms of
%the accompanying potential [5. F. Arias de Saavedra, J. Boronat, A. Polls and A. Fabrocini,
%Phys. Rev. B50, 4248 (1994); A. Puente, Ll. Serra and M. Casas, Z. Phys. D31, 283 (1994).]
%
%
In the present paper, after a clarifying discussion of the ordering problem (Sec. \ref{ordering}),
we analyze the PDM Hamiltonian for a solitonic mass profile (Sec. \ref{sectmass}).
First, we study in detail the nontrivial PDM $V=0$ case (Sec. \ref{secnull})
and thereafter add a $\tanh x$ potential function (Sec. \ref{tanh}).
This potential is related to nanowire step structures with a varying radius \cite{willatzenlassen2007} 
and hyperbolic versions of Scarf
\cite{scarf}, Rosen-Morse \cite{rosenmorse} and Manning-Rosen potentials \cite{manningrosen}
of interest in modeling molecular vibrations and intermolecular forces,  recently discussed in e.g. \cite{znojil,ozlem,souzadutra,chinos2008,chinos2009,chinos201011,correa2012,bharali2013,canadian2013}.
In Sec. \ref{secspecial} we also discuss a special case of the form $\sinh^2 x$.
In all these cases we analytically find
the complete set of solutions to the PDM Schrodinger equations
by means of a series of coordinate transformations and wave function mappings.
In Sec. \ref{conclusion} we draw our conclusions.
\section{The PDM ordering problem \label{ordering}}

We start this paper by analyzing the \textit{Hermitian} kinetic %effective
operator
\bea
& & \hat T =\,\frac 18\left\{\left(m^{-1}(\vec r)\,\hat p^2\,+\,\,\hat
p^2\,m^{-1}(\vec r)\right) +\,  m^{\alpha }(\vec{r})\,\hat{p}\ m^{\beta }(\vec{r%
})\,\hat{p}\  m^{\gamma }(\vec{r})\,\,+\,m^{\gamma }(\vec{r})\,\hat{p}
\ m^{\beta }(\vec{r})\,\hat{p}\ m^{\alpha }(\vec{r})\, \right\}\label{Top}
\eea
with the following constraint on the parameters $\alpha +\beta +\gamma =-1$
as proposed in \cite{vonroos}.
The first term, not considered originally, is here added just in order to include
the usual symmetrized or Weyl ordered operator \cite{tdlee} (viz. $\alpha \,=\,\gamma=\,0$)
 in the general expression.

In one dimension, when one properly commutes the momentum
operators $\hat p=-i\hbar \frac d{dx}$ to the right,
the following effective operator is obtained
\begin{equation}
\hat T\,=\,\frac{1}{2\,m}\,\hat{p}^{2}\,+\,\frac{i\hbar }{2}\frac{dm/dx}{m^{2}}%
\,\,\hat{p}\,\,+\,U_{\rm k}\left( x\right),
\end{equation}
where
\beq
U_{\rm k}\left( x\right) =\frac{-\hbar ^{2}}{
4m^{3}}\left[ \left( \alpha +\gamma -1\right) \frac{m}{2}\frac{
d^{2}m}{dx^{2}}+\left(1-\alpha \gamma -\alpha -\gamma
\right) \left( \frac{dm}{dx}\right) ^{2}\right]\label{kinpot}
\end{equation}
As a result of the general proposal (\ref{Top}), we obtain
an effective potential $U_{\rm k}$ of kinematic origin, which is a different
function for different combinations of the mass power parameters $\alpha, \beta, \gamma$.
This lack of uniqueness can be eliminated by imposing the condition
\begin{equation}
\alpha \,+\,\gamma \,=\,1\,=\alpha \,\gamma \,+\,\alpha\,+\,\gamma  \label{cond}
\end{equation}
which implies that for  $\alpha =0$ and $\gamma=1$, or  $\alpha=1 $ and $\gamma =0$,
the kinetic operator $\hat T$ is free of the uncertainties %ambiguities
coming from the commutation rules of Quantum Mechanics ($[\hat x,\hat p]=i\hbar$).
Note that condition (\ref{cond})  excludes the possibility of a Weyl
ordering (and consequently the kinetic operator used in \cite{likhun}; see \cite{DA}).

Even in this case, although free of the ambiguous kinematic potential (\ref{kinpot}),
the effective Schr\"odinger equation
will contain a first order derivative term. For an arbitrary external
potential $V(x)$ the non-ambiguous PDM Schr\"odinger equation results
\beq %
\left[\frac{1}{2m}\,\hat{p}^2 +\frac{i\hbar}{2}
\frac{m'}{m^2}\,\hat p + V(x)\right]\Psi (x) = E \Psi (x). \label{H} %\label{BDD}
\eeq %

\section{Smooth mass profile \label{sectmass}}

%%%%%%%%%%%%%%%%%%%%%%%%%%%%%%%%%%%%%%%%%%%%%%%%%%%%%%%%*
As we have seen, in order to avoid ambiguities from the beginning we were led to
equation (\ref{H}). It coincides with the Ben Daniel-Duke ordering of the PDM problem \cite{BDD}.
This ordering has recently been shown to be appropriate for the
understanding of growth-intended geometrical nanostructures
suffering from size variations, impurities, dislocations,
and geometry imperfections \cite{willatzenlassen2007}, and
is also consistent with the analysis made in \cite{renan} where the Dirac equation
was considered.

After substitution of the momentum operator, Eq. (\ref{H}) can be written  as
\beq %
\left \{ \frac{d^2}{dx^2} -\frac{1}{m(x)} \frac{d m(x)}{dx} \frac{d}{dx} +
\frac{2m(x)}{\hbar^2} [E-V(x)] \right \}
\psi(x) =0. \label{pdm}
\eeq %
Here we adopt the following smooth effective mass distribution
\beq m(x)=m_0 \textsf{sech}^2(a x) \label{massa} \eeq
because it is an appropriate representative of a solitonic profile
(see e.g. \cite{iran} and \cite{bagchi})
found in several effective models of condensed matter and low energy nuclear physics.
%  btw, it looks appropriate for the aimed family of potentials.
In Eq. (\ref{massa}), $a$ is a scale parameter that widens the shape of the
effective mass as it gets smaller. This is equivalent to diminish the mass and
 energy scales (see Eq. (\ref{eqSch}) ).
In this case, the effective differential equation reads
\beq %
\psi'' (x)+2 \tanh(x)
\psi'(x)+ \frac{2m_0}{a^2\hbar^2}(E-V(x)) \textsf{sech}^2(x) \psi(x)=0, \label{eqSch}%%
\eeq %%
where we shifted $a x \rightarrow x$.
By means of
\beq
\psi(x)=\cosh^{\nu}\!(x)\,\varphi(x) \label{transf} \eeq
it becomes
\bea%%
\varphi''(x)+2 (\nu+1) \tanh(x)\varphi'(x)+\left[\nu(\nu+2) \tanh^2\!x + (\nu +\frac{2m_0}{a^2\hbar^2}(E-V(x)) )\textsf{sech}^2(x) \right] \varphi(x)=0 \label{eqtransf} \eea %%
%
%Nota-se que  (\ref{transf}) preservou a paridade da Eq. (\ref{eq2}).
%
We next transform $x \rightarrow z$ by
\beq   \frac{dz}{dx}= \textsf{sech}\,x, \label{xz}\eeq
namely
\beq \cos\,z = \textsf{sech}\,x, \label{transf2} \eeq
which maps $(-\infty,\infty) \rightarrow (-\frac{\pi}{2},\frac{\pi}{2})$.

Now the equation in the  $z$ variable reads
\bea %%
\varphi''(z)+(2\nu+1)\tan(z)\varphi'(z)+\left[
\nu+\nu(\nu+2)\tan^2(z)+\frac{2m_0}{a^2\hbar^2}(E-V(z)) \right]
\varphi(z)=0 %%
\eea%%
If we choose $\nu=-1/2$ we can eliminate the first derivative, resulting in
\beq %%
-\frac{d^2 \varphi(z)}{dz^2}+ \left [ \frac{1}{2} +\frac{3}{4}\tan^2(z)+
\tilde{V}(z)\right]\varphi(z)=\mathcal{E}\varphi(z) \label{schrodinger-cons} %%
\eeq %%
where $\tilde{V}(z)=\frac{2m_0 }{a^2\hbar^2} V(z)$ and
$\mathcal{E}=\frac{2m_0}{a^2\hbar^2}E$.
This equation allows symmetric and antisymmetric solutions provided  $\tilde{V}(z)$ (and
correspondingly $V(x)$) is even.

We can see that Eq. (\ref{pdm}) is thus equivalent to a regular
\textit{constant}-mass stationary Schrodinger equation, Eq.(\ref{schrodinger-cons}),
for a particle of mass $m_0$  in a
\textit{confining} potential %(except in the particular case  $\tilde{V}(z)=-\frac{3}{4}\tan^2(z)+$cons.)
\beq %%
\mathcal{V}(z)=\frac{1}{2} +\frac{3}{4}\tan^2(z)+\tilde{V}(z). \label{mathcalV}%%
\eeq %%
Its dynamics is restricted to within
$z =(-\frac{\pi}{2},\frac{\pi}{2})$ where $\varphi(z=\pm \frac{\pi}{2})=0$.
\section{The case $V(x)=0$ \label{secnull}}

\begin{figure}[htb]
\center
\includegraphics[width=7.cm,height=5.5cm]{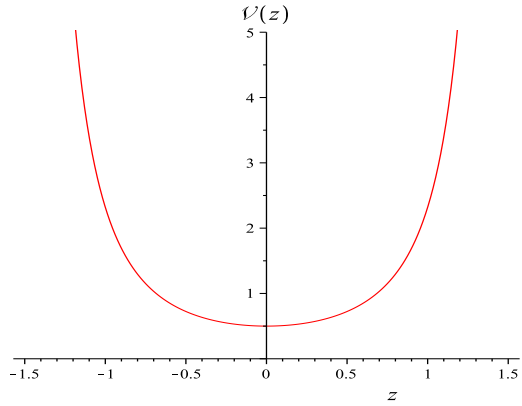}
\caption{\label{pot_z} Potential in $z$ space; $V(x)=0$ or
$\mathcal{V}(z) = \frac{1}{2}+\frac{3}{4} \tan^2(z)$.}
\end{figure}

In the $V(x)=0$ case (see Fig. \ref{pot_z}) we can analytically solve  the  Schrodinger equation (\ref{schrodinger-cons})
in terms of a new variable,  $y = \cos z$, where $0< y < 1$.
Since $E>0$, we can define $k^2=2m_0E/(a^2\hbar^2)$ and obtain
\beq \varphi''(y)+ \frac{y}{y^2-1}\varphi'(y)+
\left(\frac{\meio-k^2 }{y^2-1} -\frac{\tquarto}{y^2}\right)
\varphi (y) = 0. \label{varphi1} \eeq \noindent
Now, we define another wave function $\varphi(y)=y^{-1/2}h_1(y)$ in order to put it in a more
familiar form
\beq %%
h_1''(y)+\left( \frac{-1}{y}+\frac{\meio}{y-1}+\frac{\meio}{y+1}\right)
h_1'(y)+\frac{-k^2 y}{y (y-1) (y+1)}h_1(y) = 0\,\,
\label{eqh} %%
\eeq %%
This equation
belongs to the second order fuchsian class \cite{hille} and
can be recognized as a special case of the Heun equation \cite{ronveaux}
\beq
H''(y)+\left( \frac{\gamma}{y}+\frac{\delta}{y-1}+\frac{\varepsilon}{y-d} \right) H'(y)+\frac{\alpha \beta y-q}{y(y-1)(y-d)} H(y) =0\,\, \label{heun} \eeq
%with regular singularities at $(0, 1, d, \infty)$ ($d\neq (0,1,\infty)$).
where the parameters obey the fuchsian relation $\alpha+\beta+1=\gamma+\delta+\varepsilon$.
In the neighborhood of each singularity two linearly independent local solutions can be identified
by their characteristic exponents (two for each solution) out from the Frobenius series.
Although not so well known as its relative, the hypergeometric Gauss equation, a set of 192
different expressions have been recently given for the Heun equation by means of a set of transformations of a group of automorphysms  \cite{maier}.

%Para as solucoes L.I. pode-se utilizar o simbolo-P de Riemann dado por
%\[ P\,\left\{
%\begin{array}{ccccc}
%  0 & \,1 & d & \,\,\infty & {} \\
%  0 & \,0 & 0 &\,\, \alpha & \,\,y \\
%  1 - \gamma& \,1-\delta & \,1-\varepsilon & \,\,\beta & {}
%\end{array}
%\right\} \]

Since we want the solution to Eq.(\ref{pdm}) about $x=0$ here we will look for
solutions  around the singularity $y=1$ where the characteristic exponents are $0$ and $1-\delta$.
Therefore, the two L.I. local solutions of interest to Eq.\ref{heun} are
\bea %
 H = H\big(1-d,-q+\alpha \beta ,\alpha,\beta,\delta,\gamma; 1-y\big) \hspace{5cm}\\
 H=(1-y)^{1-\delta}\, H\big[1-d,-q+(\delta-1)\gamma d + (\alpha-\delta+1)(\beta-\delta+1),\, \beta-\delta+1, \alpha-\delta+1, 2- \delta, \gamma; 1-y\big] %
\eea %

\subsection{Even solutions}
By comparing Eq.(\ref{eqh}) and Eq.(\ref{heun}) and using the fuchsian relation above
we identify $d=-1$, $q=0$, $\alpha=-\meio +\meio\sqrt{1+4k^2}$, $\beta=-\meio
-\meio\sqrt{1+4k^2}$, $\gamma=-1$,  $\delta=\meio$ and $\varepsilon=\meio$
resulting in
\bea
h_1^{(1)}(y) = H\left(2,-k^2,\frac{-1+\sqrt{1+4k^2}}{2},
\frac{-1-\sqrt{1+4k^2}}{2},\meio,-1; 1-y\right) ~~~~\label{eqh_par}\\
 h_1^{(2)}(y) = (1\!-\!y)^{1/2}\!H\!\left(\!2,-\frac{3+4k^2}{4},
-\frac{\sqrt{1+4k^2}}{2},\frac{\sqrt{1+4k^2}}{2},\tmeio,\!-1;
1\!-\!y\right)~~
\eea
for the first of the two L.I. solutions. Recalling that $y=\cos z$
we have

\beq
\varphi^{(1,2)}(z) = \sqrt{\sec z}\,h_1^{(1,2)}(\cos z)
\eeq
for the potential $\mathcal V(z)$ [Eq. (\ref{mathcalV})] about $z=0$ .
Finally, in the original $x$ space, the even solutions of the PDM
 differential equation for the quantum particle of mass $m(x)$ [Eq. (\ref{massa})] read
\beq \psi^{(1,2)}(x) = h_1^{(1,2)}(\textsf{sech} x). \eeq
Naturally, we still need to impose the relevant boundary conditions. This is easily done in $z$
since the potential diverges to $+\infty$ in $z=\pm \frac{\pi}{2}$ which implies $\varphi(z=\pm
\frac{\pi}{2})=0$. The first solution is convergent only for $k^2=n(n+1)$ with $n=1,3,...$, and
the second solution is not acceptable because it is not differentiable at $z=0$ (see below).
\subsection{Odd solutions}

In order to obtain the odd solutions we first transform
\beq
 \varphi(z)=\sin z\, \phi(z)
 \eeq
in Eq. (\ref{schrodinger-cons}), resulting in equation
\beq \phi''(z) + 2 \cot z\,
\phi'(z)-\left(\frac{3}{2}+\frac{3}{4} \tan^2 z \right) \phi(z) = -k^2 \phi(z). \eeq

Now using again $y=\cos z$ we get
\beq
\phi''(y)+ \frac{3y}{y^2-1}\phi'(y)+\left(\frac{\tmeio-k^2 }{y^2-1}
-\frac{\tquarto}{y^2}\right) \phi (y) = 0
\label{varphi2} \eeq
and once more $\phi(y)=y^{-1/2}h_2(y)$ leads to
\beq
h_2''(y)+\left( \frac{-1}{y}+\frac{\tmeio}{y-1}+\frac{\tmeio}{y+1}\right)
h_2'(y)+\frac{-k^2 y}{y (y-1) (y+1)}h_2(y) =0\,\, \label{eqh2}
\eeq %%
As before $h_2(y)$ are local Heun functions around $y=1$ (namely $z=0$ and $x=0$)) given by
\bea
 h_2^{(1)}(y) = H\left(2,-k^2,\frac{1+\sqrt{1+4k^2}}{2},
\frac{1-\sqrt{1+4k^2}}{2},\frac{3}{2},-1; 1-y\right) ~~~~~\label{eqh_impar}\\
 h_2^{(2)}(y) = (1\!-\!y)^{-1/2} H\!\left(\!2,\quarto- k^2,
-\frac{\sqrt{1+4k^2}}{2}, \frac{\sqrt{1+4k^2}}{2}, \frac{1}{2},-1;
1-y \right) ~~~
\eea
with the following identification of parameters
 $\alpha_2=\frac{1+\sqrt{1+4k^2}}{2}$, $\beta_2=\frac{1-\sqrt{1+4k^2}}{2}$,
$\gamma_2=-1$ and $\delta_2 = \varepsilon_2 = \frac{3}{2}$.
These solutions,  in $z$ and $x$ space respectively, read
\bea
\varphi^{(1,2)}(z) \!\!&=&\!\! \sin\! z(\sec z)^{1/2}~h_2^{(1,2)}(\cos z)\\
\psi^{(1,2)}(x) \!\!&=&\!\! b \tanh x~h_2^{(1,2)}(\textsf{\textsf{sech}} x)~~~~~
\eea%
Here, the first solutions, $\varphi^{(1)}(z)$ and $\psi^{(1)}(x)$, converge only
for $k^2=n(n+1)$, with $n>0$ even. Again, $\varphi^{(2)}(z)$ ($\psi^{(2)}(x)$)
is not differentiable at the origin and we discard it.

\subsection{Hypergeometric solutions}

In several cases, the Heun equation can be reduced to a ordinary hypergeometric equation.
This is actually the present case, where we can see that the solutions are products of
trigonometric functions and hypergeometric ones.

Indeed, Maier \cite{maier2005} has recently determined
that for a nontrivial Heun equation ($\alpha \beta \neq 0$ or $q \neq 0$) its local
 solution $H(d, q, \alpha, \beta, \gamma, \delta; t)$ can be reduced to
${}_2F_1(a,b,c; R(t))$ where $a,b,c$ depend on $d, q, \alpha, \beta, \gamma$ and
$R(t)$ is a polynomials of up to sixth order provided $q=\alpha \beta p$, where $(d,p)$
is one among a set of 23 different values.

In particular, the case
($d,p$)=($2,1$) allows a transformation $H(t) \rightarrow {}_2F_1(R(t))$
with $R(t)$ of order 2 or 4.
Now the even and odd physically acceptable solutions [Eqs. (\ref{eqh_par}) and (\ref{eqh_impar})]
can be written as
\bea
h_{\rm phys}^{(1)}(t) = H\left(2,\alpha_1 \beta_1,\alpha_1,\beta_1,\gamma_1',\delta_1'; t\right)\\
h_{\rm phys}^{(2)}(t) = H\left(2,\alpha_2 \beta_2, \alpha_2,\beta_2,\gamma_2',\delta_2'; t \right) \eea
where $\alpha_{1,2}$ and $\beta_{1,2}$ are those defined in the previous section
and $\gamma_1'=1/2$, $\delta_1'=-1$, $\gamma_2'=\tmeio$, $\delta_2'=-1$ and $t=1-y$.

The equations above are Heun solutions  corresponding to $(d,p)=(2,1)$ with redefined parameters.
Thus, we can simply reduce them to hypergeometric expressions
\bea
h_{\rm phys}^{(1)}(t) = H\left(2,\alpha_1 \beta_1,\alpha_1,\beta_1,\gamma_1',\delta_1'; t\right) =\!{}_2F_1\left(\frac{\alpha_1}{2}, \frac{\beta_1}{2}, \gamma_1'; t(2-t) \right)\\
h_{\rm phys}^{(2)}(t) = H\left(2,\alpha_2 \beta_2, \alpha_2,\beta_2,\gamma_2',\delta_2'; t \right)
=\!{}_2F_1\left(\frac{\alpha_2}{2}, \frac{\beta_2}{2}, \gamma_2'; t(2-t) \right) \eea

Then, in terms of the space variables $z$ and $x$ the following
\bea
\varphi_{\rm phys}^{(1)}(z) = \sec^{\!1/2}\!z\ {}_2F_1\left(\frac{-1+\sqrt{1+4k^2}}{4},
\frac{-1-\sqrt{1+4k^2}}{4}, \frac{1}{2}~ ;\, \sin^2\!z \right) \label{fis_sim_z}\\
 \varphi_{\rm phys}^{(2)}(z) = \sin\!z\sec^{\!1/2}\!z\ {}_2F_1\left(\frac{1+\sqrt{1+4k^2}}{4}, \frac{1-\sqrt{1+4k^2}}{4},
\frac{3}{2} \,;\, \sin^2\!z \right) \label{fis_asim_z} \eea
and
\bea
\psi_{\rm phys}^{(1)}(x) = {}_2F_1\!\left(\frac{-1+\sqrt{1+4k^2}}{4},
\frac{-1-\sqrt{1+4k^2}}{4}, \frac{1}{2};\, \tanh^2\!x \right) \label{fis_sim_x}~~~~\\
\psi_{\rm phys}^{(2)}(x) = \tanh\!x\ {}_2F_1\!\left(\frac{1+\sqrt{1+4k^2}}{4},
\frac{1-\sqrt{1+4k^2}}{4}, \frac{3}{2} ; \tanh^2\!x \right)\,\, \label{fis_asim_x} \eea
are, respectively,
the physically acceptable solutions to the equation.

It can be shown that $\psi_{\rm phys}^{(1,2)}$ are mutually orthogonal \beq \langle
\psi_{n}^{(i)}, \psi_{m}^{(j)}\rangle = \delta_{nm} \delta_{ij}, \eeq
and generate a complete set of solutions to the problem.
Here $i, j =1, 2$  and $n, m$ represent the order of the solution.

\subsection{Conditions of existence}

In Eq. (\ref{fis_sim_z}), since $\lim_{(z \rightarrow \pm \frac{\pi}{2})}\ \sec\!z = \infty$,
 we must impose
\beq \lim_{z \rightarrow \pm
\frac{\pi}{2}} {}_2F_1\left(\frac{-1+\sqrt{1+4k^2}}{4}, \frac{-1-\sqrt{1+4k^2}}{4},
\frac{1}{2}~ ;\, \sin(z)^2 \right)=0. \eeq
Since, for this expression
\beq \lim_{z \rightarrow \pm \frac{\pi}{2}} {}_2F_1\left(z \right)
\rightarrow \frac{\sqrt{\pi}}{\Gamma \left(\frac{3}{4}-\frac{1}{4}
\sqrt{1+4k^2} \right) \Gamma \left(\frac{3}{4}+\frac{1}{4}
\sqrt{1+4k^2}\right)} \eeq
the vanishing condition results
\beq
\frac{3}{4}-\frac{1}{4} \sqrt{1+4k^2} = -p, ~~ p =0, 1, ... \eeq
for $k$ finite. Regrouping, we obtain
\beq k^2 = (2p+1)^2+(2p+1). \label{odd}\eeq

On the same token, for the antisymmetric solutions, Eq.(\ref{fis_asim_z}), we need
\beq \lim_{z \rightarrow \pm
\frac{\pi}{2}} {}_2F_1\left(\frac{1+\sqrt{1+4k^2}}{4}, \frac{1-\sqrt{1+4k^2}}{4},
\frac{3}{2}~ ;\, \sin^2\!z \right) =0. \eeq
Since, in this case,
 \beq \lim_{z \rightarrow \pm \frac{\pi}{2}} {}_2F_1\left(z \right)
\rightarrow\frac{2\sqrt{\pi}}{k^2\Gamma \left(\frac{1}{4}-\frac{1}{4}
\sqrt{1+4k^2} \right) \Gamma \left(\frac{1}{4}+\frac{1}{4}
\sqrt{1+4k^2}\right)} \eeq we need
 \beq \frac{1}{4}-\frac{1}{4} \sqrt{1+4k^2} = -q, ~~ q =0, 1, ... \eeq
which, after regrouping may be written as
\beq
k^2=2q (2q+1). \label{even}
\eeq
Substituting  $2p+1$ (2q) by $n$ in Eq. (\ref{odd}) (Eq. \ref{even}),
the existence condition is in both cases $k^2= n(n+1)$, with $n$ odd (even) for even (odd)
solutions, respectively.

Note that this result determines that energy is quantized   as
\beq %
E = \frac{a^2\hbar}{2m_0}~ n(n+1),
\eeq %
with $n>0$, since energy cannot be zero.

The final expressions for symmetric and antisymmetric solutions (in $z$ and $x$ spaces) are thus
\bea
\varphi_{\rm phys}^{(1)}(z) \!\!&=&\!\! \sec^{\!1/2}\!z \ {}_2F_1\left(\frac{n}{2},
-\frac{n+1}{2}, \frac{1}{2}~ ;\, \sin^2\!z \right) \label{fis_sim_z_n}\\
\varphi_{\rm phys}^{(2)}(z) \!\!&=&\!\! \sin\!z\sec^{\!1/2}\!z\
{}_2F_1\left(\frac{n+1}{2}, -\frac{n}{2}, \frac{3}{2} \,;\, \sin^2\!z
\right) \label{fis_asim_z_n} \eea
and
\bea
\psi_{\rm phys}^{(1)}(x) \!\!&=&\!\! {}_2F_1\!\left(\frac{n}{2},
-\frac{n+1}{2}, \frac{1}{2}~ ;\, \tanh^2\!x  \right) \label{fis_sim_x_n}~~~~\\
\psi_{\rm phys}^{(2)}(x) \!\!&=&\!\! \tanh\!x \ {}_2F_1\!\left(\frac{n+1}{2},
-\frac{n}{2}, \frac{3}{2} \,;\, \tanh^2\!x \right).
\label{fis_asim_x_n} %%
\eea

\begin{figure}[htb]
 \center
\includegraphics[width=7.cm,height=5.5cm]{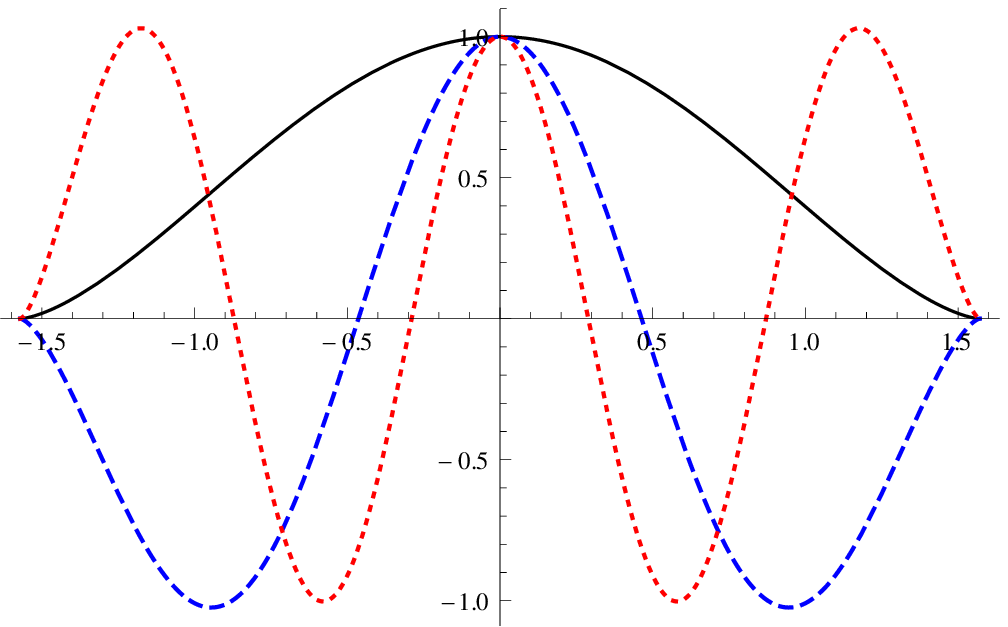}
\caption{\label{sim_z} Symmetric solutions in $z$ space, Eq. (\ref{fis_sim_z_n}),  for $n = 1$
(solid line),  $n = 3$ (dashed line)  and $n = 5$ (dotted line).}
 \end{figure}

\begin{figure}[htb]
\center %
%{\includegraphics[width=7.0cm,height=5.5cm]{sim_x.pdf}}
{\includegraphics[width=7.0cm,height=5.5cm]{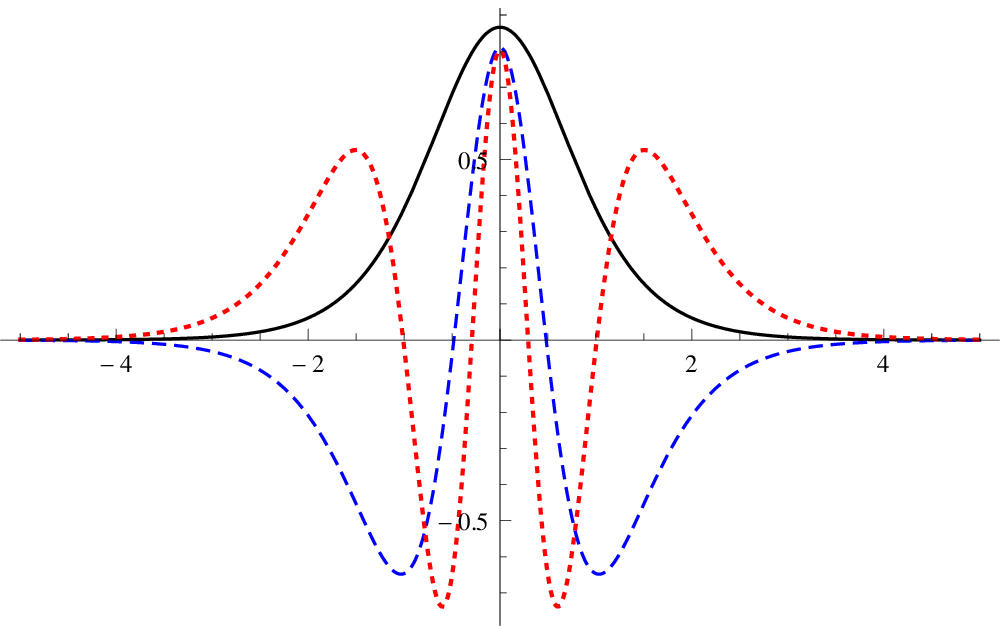}} \caption{\label{sim_x}
Symmetric normalized solutions in $x$ space ,
Eq. (\ref{fis_sim_x_n}) (with $a$=1), for $n = 1$  (solid line),  $n = 3$ (dashed line)  and $n = 5$ (dotted line).}
\end{figure}

\begin{figure}[htb]
\center %
%{ \includegraphics[width=7.cm,height=5.5cm]{asim_z.pdf}}
{ \includegraphics[width=7.cm,height=5.5cm]{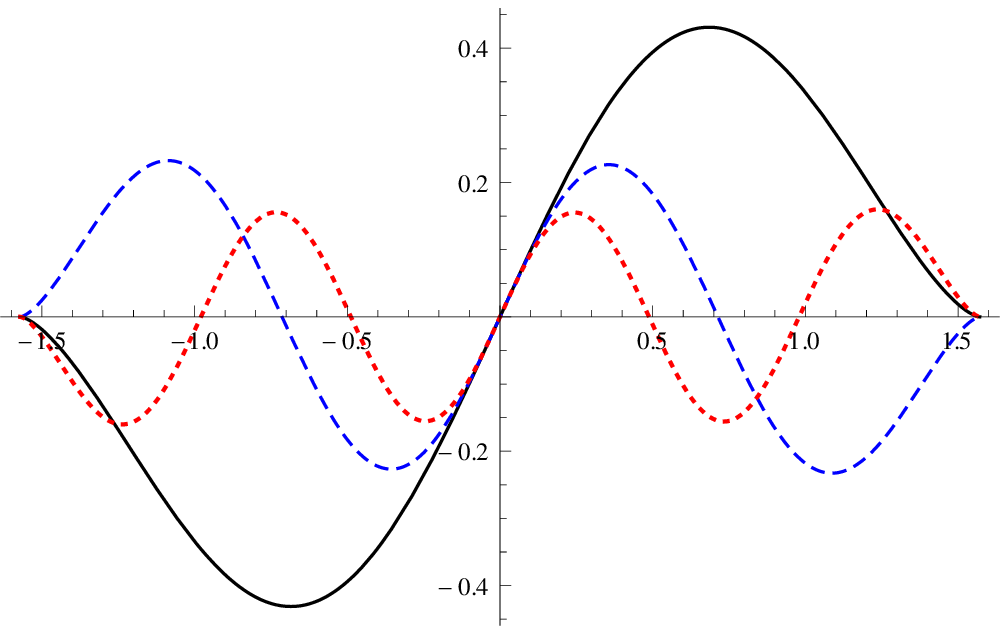}} \caption{\label{asim_z} Antisymmetric solutions in $z$ space, Eq.
(\ref{fis_asim_z_n}),  for  $n = 2$ (solid line), $n = 4$ (dashed line) and $n = 6$ (dotted line).}
\end{figure}

\begin{figure}[h!]
\center
{\includegraphics[width=7.cm,height=5.5cm]{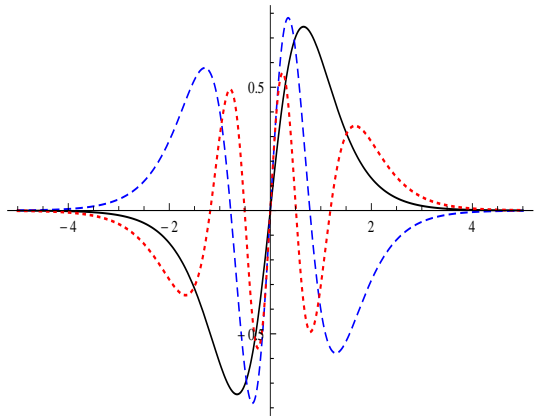}}
\caption{\label{asim_x} Antisymmetric normalized solutions in $x$ space, Eq.  (\ref{fis_asim_x_n})
  (with $a$=1), for  $n = 2$ (solid line), $n = 4$ (dashed line) and $n = 6$ (dotted line).}
\end{figure}

We illustrate Eqs. (\ref{fis_sim_z_n}) to (\ref{fis_asim_x_n}) in Figs. \ref{sim_z}, \ref{sim_x}, \ref{asim_z} and \ref{asim_x}.

In $x$ space we can get physical information about the position dependent mass particle.
Figs. \ref{sim_x} and \ref{asim_x} show the increasing probability space
density of the states around the origin.

\section{A special $V(x)= \sinh^2 x$ case \label{secspecial}}
It is interesting to note that in $z$ space, the confining problem defined by
 Eq. (\ref{schrodinger-cons})
becomes trivial for the potential $V(z)=-{3a^2\hbar^2}\tan^2z/{8m_0} + $cons.
This corresponds to a $x$ space potential function
\beq V(x)= -\frac{3 a^2\hbar^2}{8m_0}\sinh^2 x\,+\rm cons.  \label{special}\eeq
in Eq. (\ref{eqSch}).

Although this sets a nontrivial differential equation,
the related full effective potential $\cal V$(z),
Eq. (\ref{mathcalV}), is just constant and therefore
the exact solutions to the PDM problem can be easily
obtained. Interestingly, this would be a much more difficult task in the constant-mass case.

\begin{figure}[ht]
%\center %
\subfigure[]{\includegraphics[width=5.9cm,height=5.6cm]{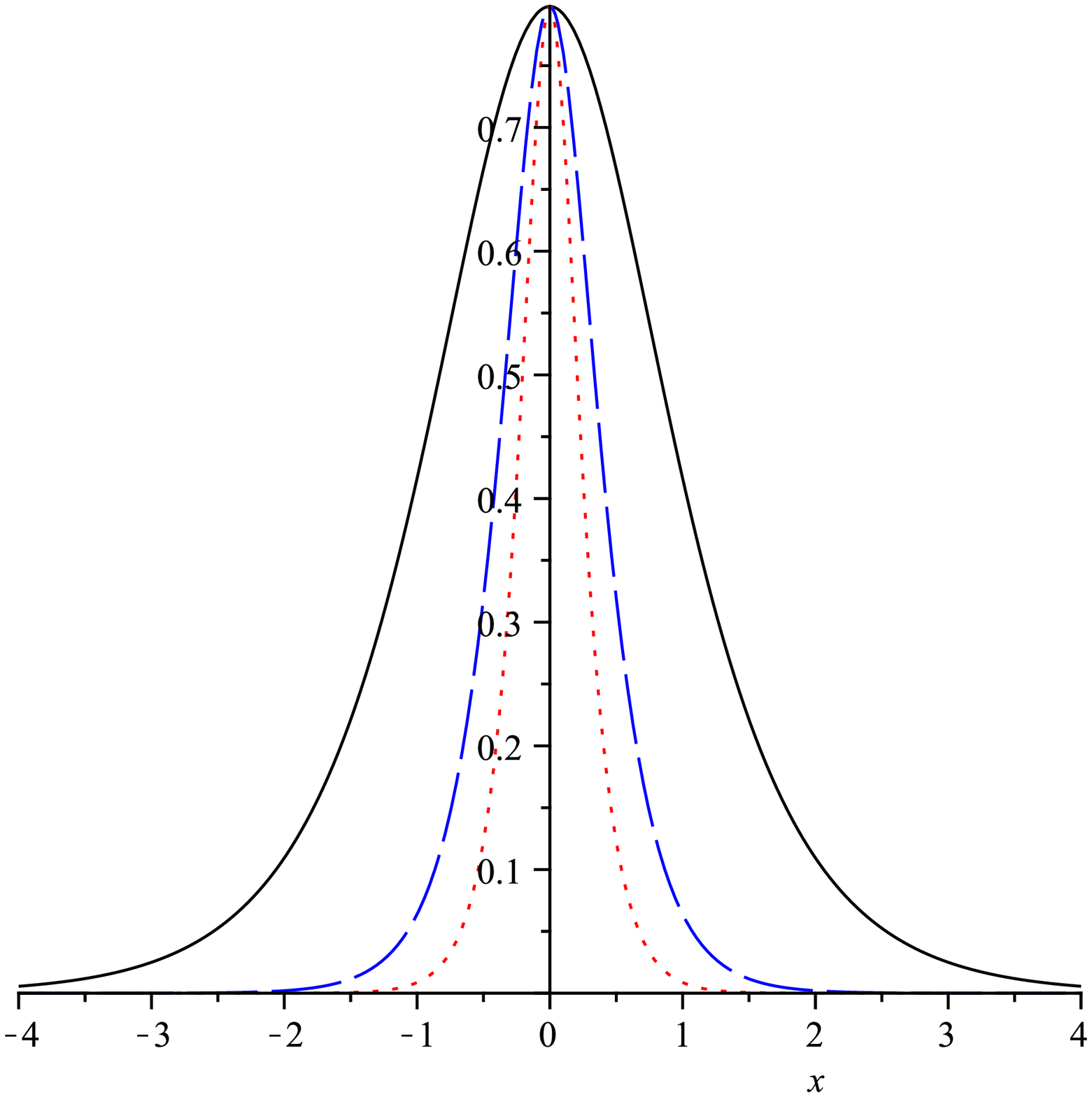}}
\subfigure[]{\includegraphics[width=5.9cm,height=5.6cm]{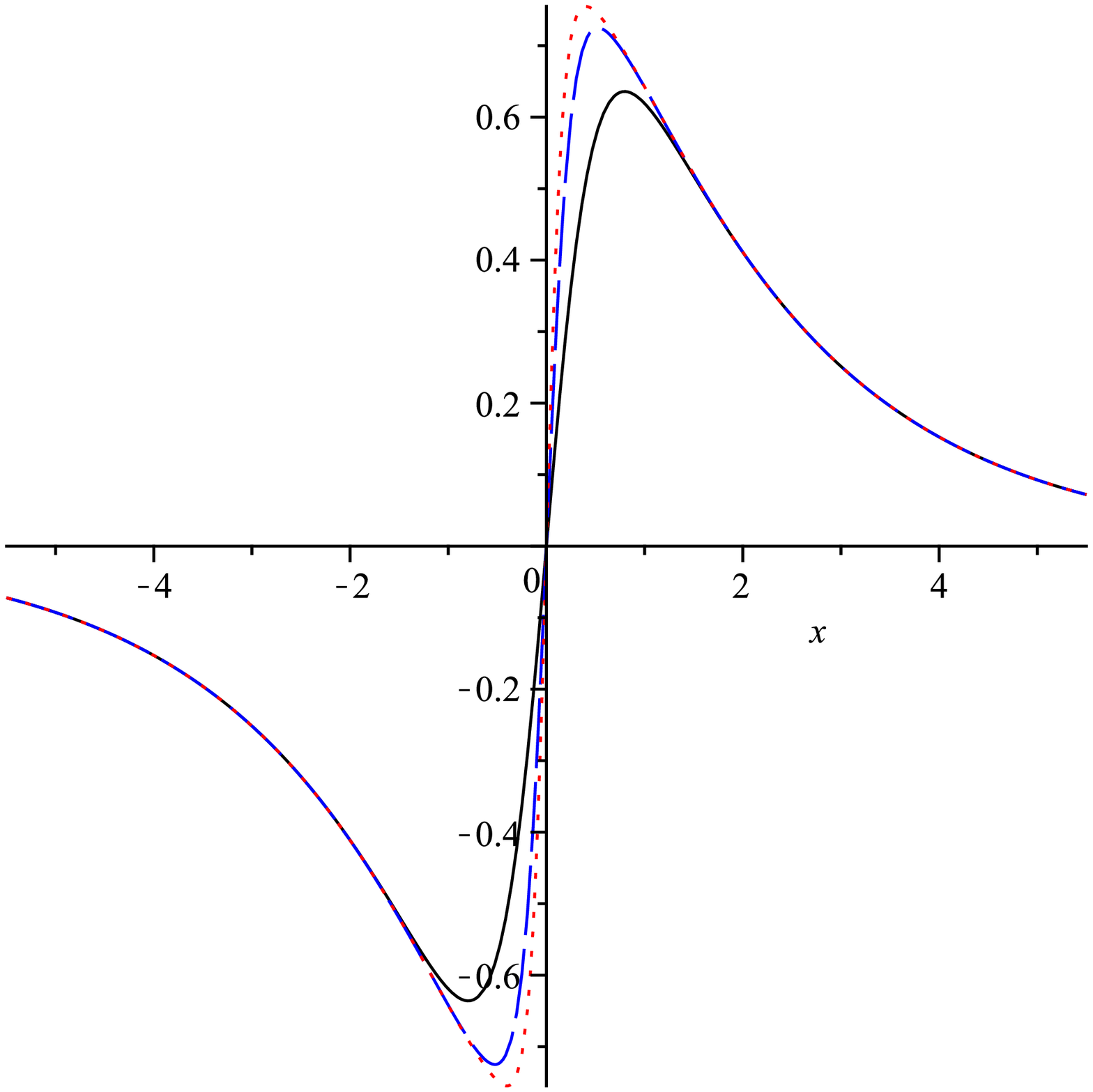}}
\caption{(a) Symmetric solutions [Eq.(\ref{sim_sinh2x})] and (b)
antisymmetric solutions [Eq. (\ref{asim_sinh2x})] for $n=1$ (solid
black line),  $n=2$ (dashed blue line), and $n=3$ (dotted red
line).\label{graf_sinh2x}}
\end{figure}

The two L.I. solutions in $z$ space read
\bea
\varphi^{(1)}(z)\!\!&=&\!\!\sqrt{\frac{2}{\pi}}\cos\left((2n+1)\,z\right]\\
\varphi(z)\!\!&=&\!\!\sqrt{\frac{2}{\pi}}\sin\left(2n\,z\right)
\eea
where we have chosen cons.=$-a^2\hbar^2/4m_0$ in Eq. (\ref{special}).
Making them vanish at the border $z=\pm \pi/2$
quantizes the energy by
\beq \mathcal{E}=\frac{a^2\hbar^2 \pi^2}{8m_0
}n^2\eeq where $n\in \mathbb{N}$. Recalling  Eq. (\ref{transf}), in $x$ space we obtain
\bea
\psi^{(1)}(x)\!\!&=&\!\!C_1 \sqrt{\frac{2}{\pi}}\textsf{sech}^{1/2} \!x \,\textsf{sech}
\big[(2n+1)\,x\big]\label{sim_sinh2x}\\
\psi^{(2)}(x) \!\!&=&\!\! C_2 \sqrt{\frac{2}{\pi}}\textsf{sech}^{1/2} \!x \,\tanh
(2n\,x),\label{asim_sinh2x}
\eea
see Fig. \ref{graf_sinh2x}.

\section{The tanh(x) case \label{tanh}}
In order to solve the associated
 differential equation we return to Eq. (\ref{schrodinger-cons}). According to
 Eq. (\ref{transf2}) in the PDM problem the potential $V(x) = V_0\tanh\!x$ corresponds to
$\tilde{V}(z)=\mathcal{V}_0 \sin(z)$, where $\mathcal{V}_0 = 2m_0V_0/(a^2\hbar^2)$.
Now, the analysis of section \ref{secnull}
implies that the PDM Schr\"odinger equation can be transformed into
\begin{equation} %%
-\frac{d^2 \varphi(z)}{dz^2}+ \mathcal{V}(z) \varphi(z)=\mathcal{E}\varphi(z)
\eeq
with
\beq
\mathcal{V}(z)= \frac{1}{2} +\frac{3}{4}\tan^2(z)+ \mathcal{V}_0\sin(z), \label{sinz}%
\eeq
See Fig. \ref{graf_pot_tg_senz}. By means of the ansatz
\beq \varphi(z)=\sqrt{\sec z}\ h(z)\eeq
  the equation above can be written as
 \beq
 h''(z) + \tan \!z \,h'(z) + (\mathcal{E}-\mathcal{V}_0 \sin\!z)\, h(z)=0.
 \eeq
\begin{figure}[ht]
 \center
{\includegraphics[width=7.cm,height=5.5cm]{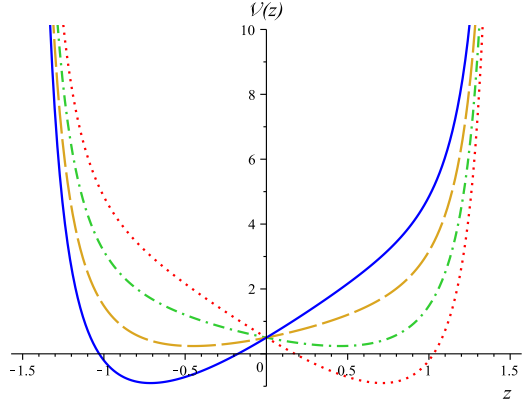}}
\caption{\label{graf_pot_tg_senz} The effective potential $\mathcal{V}(z)$, Eq. (\ref{sinz}), when  $V(x)=V_0 \tanh\!x$,
for $\mathcal{V}_0=3$ (solid blue line),  $\mathcal{V}_0 = 1$ (dashed gold line), $\mathcal{V}_0 =-1$ (dash-dotted green line) and $\mathcal{V}_0 = -3$ (dotted red line). }
 \end{figure}
With a transformation of coordinates given by
\beq y=\meio+\meio \sin \!z,\eeq
 we obtain
 \beq
 h''(y)+\left[ \frac{-(\mathcal{V}_0+\mathcal{E})+2\mathcal{V}_0 y}{y (y-1)} \right] h(y) = 0. \label{conf}
 \eeq
Equation (\ref{conf}) is thus a particular case of the \emph{confluent} Heun equation  \cite{ronveaux,hounkonnou,fiziev}
 \bea
Hc''(y) \!+\! \left( \alpha \!+\! \frac{\beta+1}{y}+\frac{\gamma+1}{y-1} \right) Hc'(y)
  \!+\!\left[ \left(\delta+\frac{\alpha}{2} (\beta+\gamma+2)\right) y \!+\! \eta+\frac{\beta}{2}+\frac{1}{2} (\gamma\!-\!\alpha) (\beta+1)\right] \frac{1}{y (y-1)} Hc(y) = 0~~ \label{confH}
 \eea
By identifying
$ \alpha=0, \,\, \beta=\gamma=-1, \,\, \delta=2\mathcal{V}_0, \,\, \eta=\meio-\mathcal{V}_0-\mathcal{E},
$
the local solutions of Eq. (\ref{conf}) about $y=0$ are given by
\bea
h^{(1)}(y) \!\!&=&\!\! y^{-\beta} Hc(\alpha,- \beta, \gamma, \delta, \eta; y) \\
h^{(2)}(y) \!\!&=&\!\! Hc^\dagger (\alpha, \beta, \gamma, \delta, \eta; y),
\eea  %
where $Hc^\dagger (y)$ is the second independent solution, so-called
 {concomitant confluent} Heun function used when $\beta=-1$ \cite{fizievCQG}.
 Since this solution diverges logarithmically when $y \rightarrow 0$
the only physically acceptable solutions are thus
\beq
\varphi(z)=\frac{\meio+\meio\sin(z)}{\sqrt{\cos(z)}}\,Hc(0,1,-1,2\mathcal{V}_0,\meio-\mathcal{V}_0-\mathcal{E}; \meio+\meio \sin{z}) \label{sol_hcz1}\,\,
\eeq
with $\varphi(z\rightarrow \pm \pi/2)=0$ for the allowed $\mathcal{E}$ and $\mathcal{V}_0$ as imposed
by the boundary conditions. Parity is not a defined symmetry in this expression for any eigenvalue. However, as we see in Fig. \ref{solz_tanh} and Fig. \ref{solx_tanh} some eigenfunctions are (a) quasi-symmetric while others are (b) quasi-antisymmetric. Note that they alternate each other, as expected.
In the original variable $x$ (recall eqs. (\ref{transf2}) and (\ref{transf})) these solutions,
which we plot in Fig. \ref{solx_tanh}, read
\beq
\psi(x)=\left({\meio+\meio\tanh{x}}\right)\,Hc(0,1,-1,2\mathcal{V}_0,\meio-\mathcal{V}_0-\mathcal{E};
\meio+\meio \tanh{x})\,\, \label{sol_hcx1}
\eeq

The energy eigenvalues can be numerically computed by imposing appropriate boundary conditions, namely,
 $\psi(x \rightarrow + \infty)=0$, ie. $\psi(y=1)=0$, with $y=\meio+\meio \tanh\!x.$
A Frobenius expansion for Eq. (\ref{sol_hcx1}) about $y=0$
\beq \psi(y)=\sum_{n=0}^\infty c_n y^n \label{psi_y} \eeq
allows this calculation.
For simplicity we choose $\mathcal{V}_0=1$ for which we obtain the list of eigenvalues presented
in Table \ref{tab:1} (for $n$ up to 25).

% For tables use
\begin{table}
% table caption is above the table
\caption{Energy eigenvalues of the Schrodinger equation for potential (\ref{sinz}).}
\label{tab:1}       % Give a unique label
% For LaTeX tables use
\begin{tabular}{llllll}
\hline\noalign{\smallskip}
$~~~~\mathcal{E}_1$ & $~~~~\mathcal{E}_2$ & $~~~~\mathcal{E}_3$ & $~~~~\mathcal{E}_4$ & $~~~~\mathcal{E}_5$ & $~~~~\mathcal{E}_6$\\
\noalign{\smallskip}\hline\noalign{\smallskip}
$1.9503339$ & $6.0115779$ & $12.0083261$ & $20.0055193$ & $30.0038467$ & $42.0028139$\\
\noalign{\smallskip}\hline
\end{tabular}
\end{table}

% Expansão em série de psi(y) em torno de y=0
%\bea
%\psi(y) &=& y- (\meio+\meio \mathcal{E})y^2+(0.2499999999+0.08333333330 \,\mathcal{E}^2)y^3 \nonumber \\
%&+&(0.02083333325-0.1041666667 \, \mathcal{E} + 0.03472222220 \, \mathcal{E}^2 \nonumber \\
%&-& 0.0069444444412\, \mathcal{E}^3)y^4+(0.03645833328-0.05833333335\,\mathcal{E} \nonumber\\
%&+&0.03263888887 \, \mathcal{E}^2-0.005555555552 \mathcal{E}^3+0.0003472222220\,\mathcal{E}^4)y^5 \nonumber \\
%&+&(0.02447916663-0.04510416669 \,\mathcal{E}+0.02493055555 \, \mathcal{E}^2 \nonumber \\
%&-&0.005069444442 \, \mathcal{E}^3+0.0004050925924 \,\mathcal{E}^4-0.00001157407407\,\mathcal{E}^5)y^6 +\dots\nonumber
%\eea
%
%
\begin{figure}[h]
 \center
 \subfigure[]{\includegraphics[width=7cm,height=6cm]{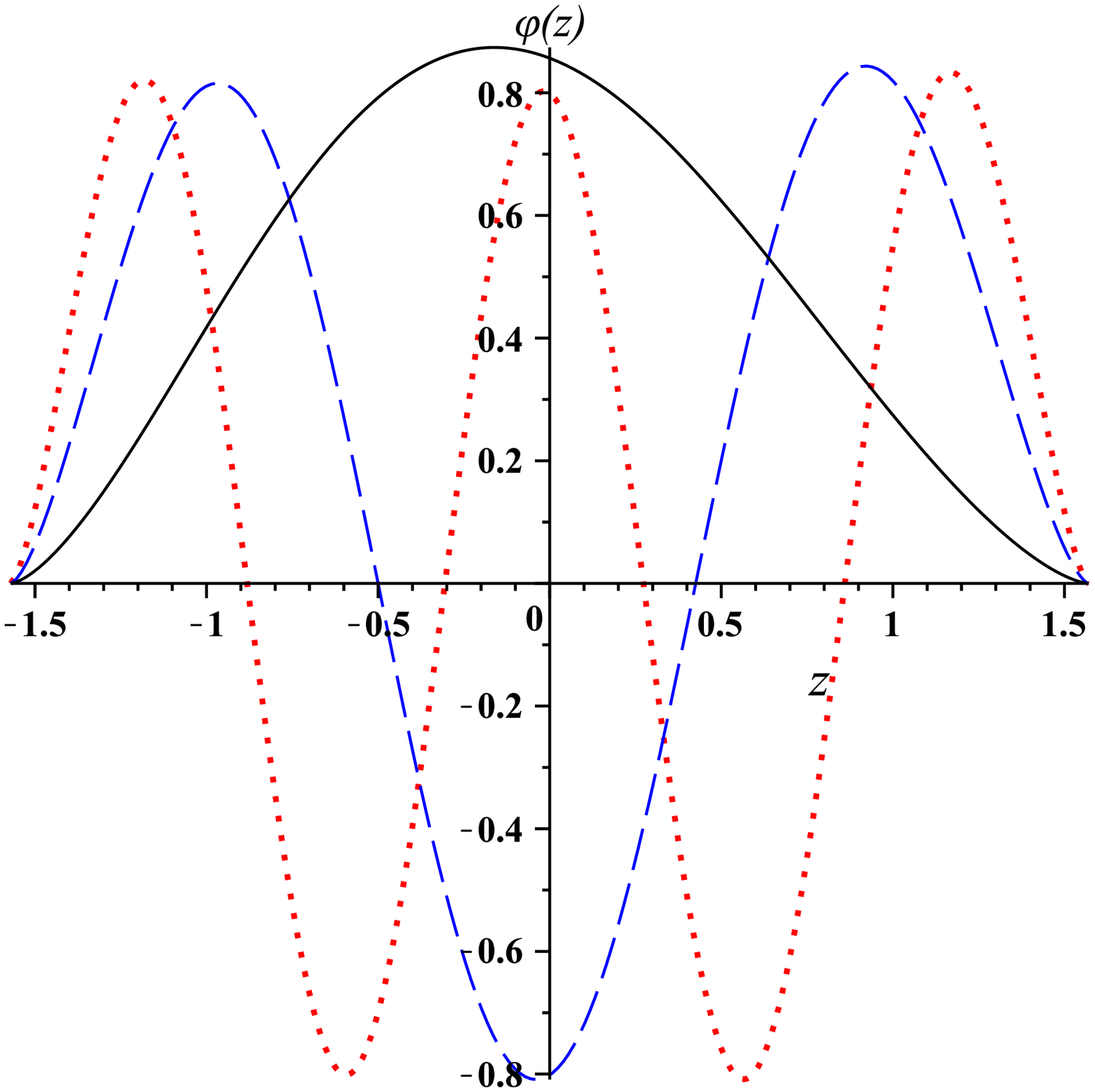}}
 \subfigure[]{\includegraphics[width=7cm,height=6cm]{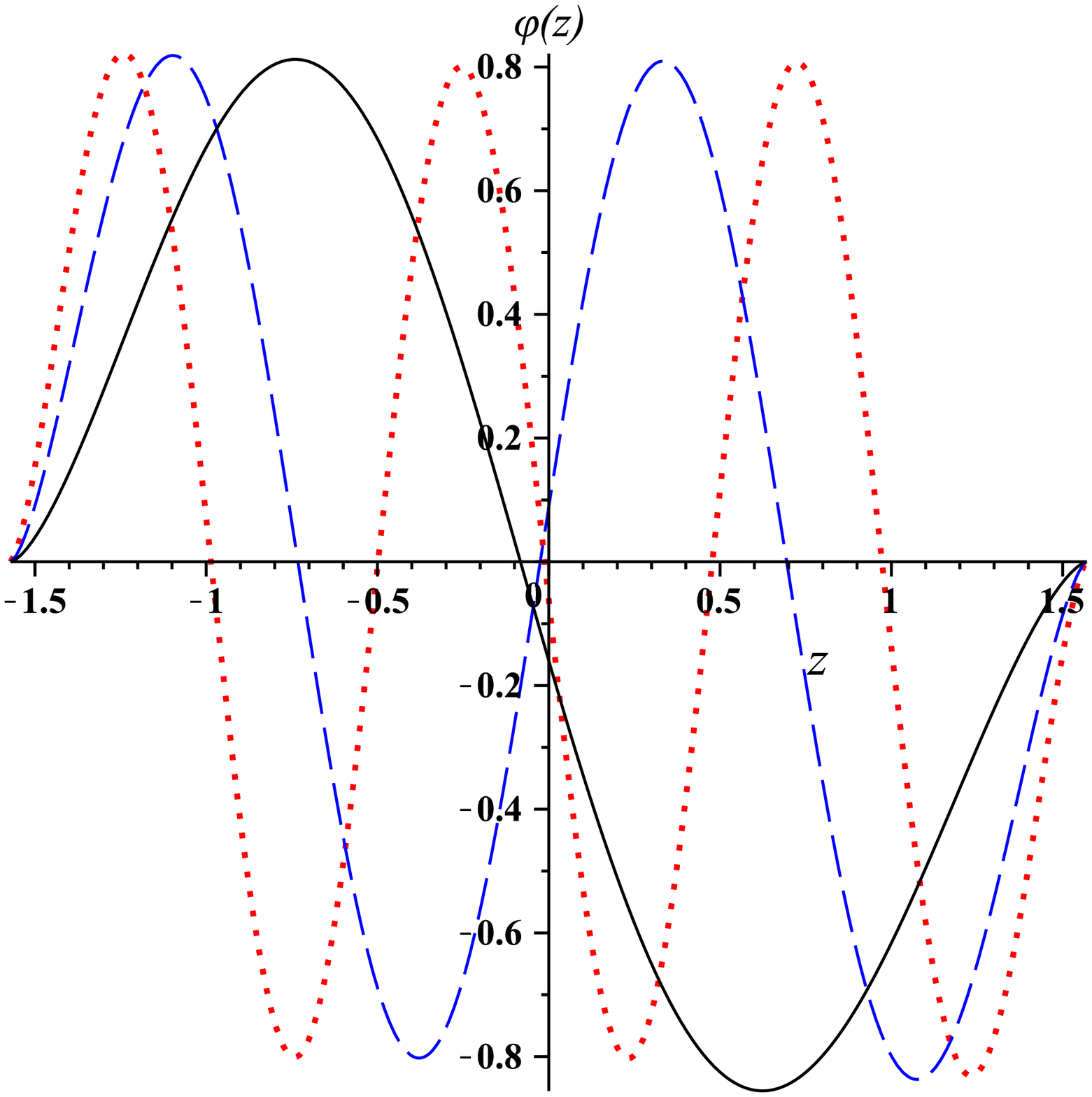}}
\caption{\label{solz_tanh} {Normalized solutions, Eq. (\ref{sol_hcz1}) with $\mathcal{V}_0=1$,
 for (a) $\mathcal{E}_1 $ (solid black line),  $\mathcal{E}_3 $ (dashed blue line),
 and $\mathcal{E}_5 $ (dotted red line), and (b) $\mathcal{E}_2 $ (solid black line),
 $\mathcal{E}_4 $ (dashed blue line) and $\mathcal{E}_6$ (dotted red line).}}
 \end{figure}
%= 1.95033397,= 6.0115779, 12.0083301,= 20.005519,=30.00384769,=42.00281609
%
\begin{figure}[h]
 \center
 \subfigure[]{\includegraphics[width=7cm,height=6cm]{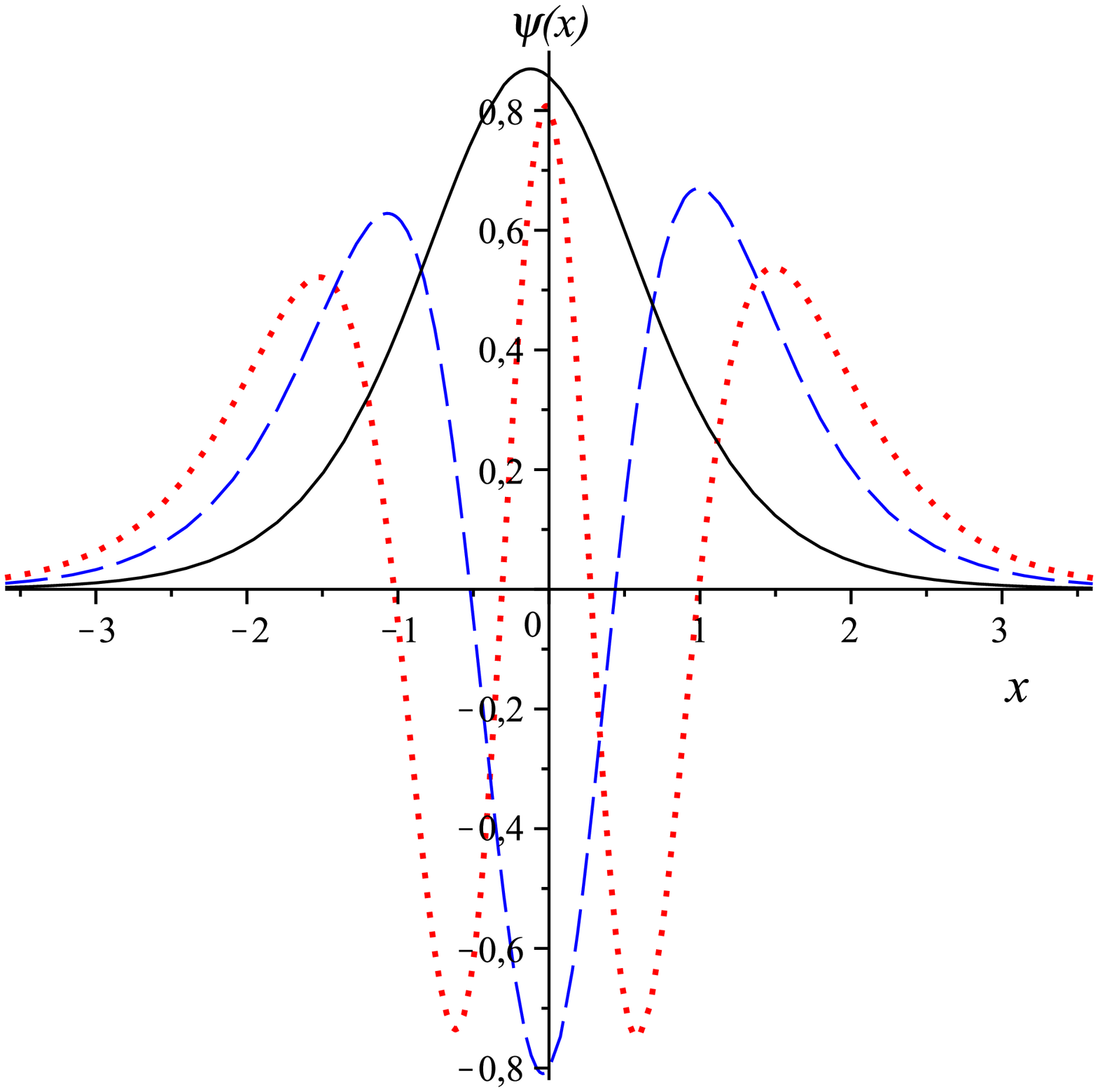}}
 \subfigure[]{\includegraphics[width=7cm,height=6cm]{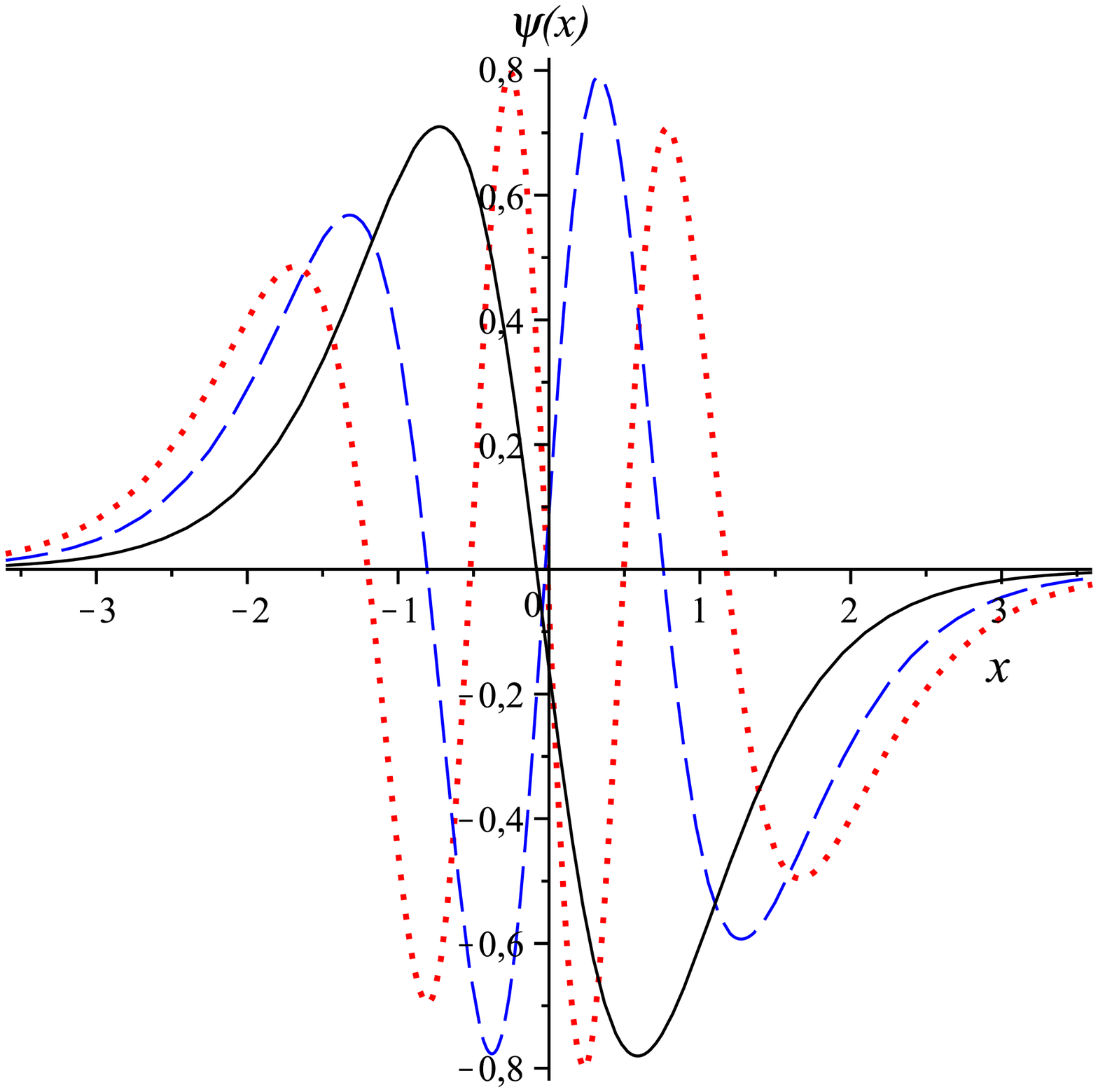}}
\caption{\label{solx_tanh} {Normalized solutions, Eq. (\ref{sol_hcx1}) with $\mathcal{V}_0=1$,
for (a) $\mathcal{E}_1 $ (solid black line),  $\mathcal{E}_3$ (dashed blue line),
and $\mathcal{E}_5 $ (dotted red line), and (b) $\mathcal{E}_2 $ (solid black line),
$\mathcal{E}_4 $ (dashed blue line) and $\mathcal{E}_6$ (dotted red line).}}
\end{figure}

\newpage
\section{Conclusion \label{conclusion}}
In this paper we have analyzed the Schrodinger equation for a nonuniform massive
particle with a solitonic mass distribution.
We have found the space of solutions related to a PDM Hermitian Hamiltonian defined by a non-ambiguous
kinetic operator and an external potential. We have shown that while a special $\sinh^2 x$
potential is easily worked out in this particular context the $V(x)=0$ case can be much more involved.
The PDM $\tanh x$ potential case can be transformed into a Heun equation which we solved exactly
by means of an analytic procedure. This potential is related to hyperbolic potentials
of special interest for modeling atomic and molecular physics.
Interestingly enough, for a long time absent in the literature,  Heun functions have recently been found
in very different contexts, see e.g. \cite{christiansencunha2011,christiansencunha2012,rumania,bulgaria,cvetic2011,herzog}.
Besides exactly obtaining all the  solutions in the three cases studied
we have plotted all the first eigenstates in a systematic way,
emphasizing their parity properties. We hope to report on further results in a forthcoming paper.

\end{document}